# Tunable exciton-optomechanical coupling in suspended monlayer MoSe$_2$

Hongchao Xie[1,2], Shengwei Jiang[1], Daniel A. Rhodes[3], James C. Hone[3], Jie Shan[1,4*], and Kin Fai Mak[1,4*]

[1] Laboratory of Atomic and Solid State Physics and School of Applied and Engineering Physics, Cornell University, Ithaca, New York, USA
[2] Department of Physics, Penn State University, University Park, Pennsylvania, USA
[3] Department of Mechanical Engineering, Columbia University, New York, USA
[4] Kavli Institute at Cornell for Nanoscale Science, Ithaca, New York, USA

*E-mails: jie.shan@cornell.edu; kinfai.mak@cornell.edu

**Abstract**
The strong excitonic effect in monolayer transition metal dichalcogenide (TMD) semiconductors has enabled many fascinating light-matter interaction phenomena. Examples include strongly coupled exciton-polaritons and nearly perfect atomic monolayer mirrors. The strong light-matter interaction also opens the door for dynamical control of mechanical motion through the exciton resonance of monolayer TMDs. Here we report the observation of exciton-optomechanical coupling in a suspended monolayer MoSe$_2$ mechanical resonator. By moderate optical pumping near the MoSe$_2$ exciton resonance, we have observed optical damping and anti-damping of mechanical vibrations as well as the optical spring effect. The exciton-optomechanical coupling strength is also gate-tunable. Our observations can be understood in a model based on photothermal backaction and gate-induced mirror symmetry breaking in the device structure. The observation of gate-tunable exciton-optomechanical coupling in a monolayer semiconductor may find applications in nanoelectromechanical systems (NEMS) and in exciton-optomechanics.

Atomically thin materials have attracted tremendous interests in the studies of NEMS devices because of their extreme lightweight, mechanical flexibility [1,2], gate-tuneability [3,4,5] and strong mechanical nonlinearity [4,5,6,7]. They also hold great promise in many sensing applications [8,9,10,11]. Although the mechanical aspects of NEMS devices based on these materials have been extensively studied in the literature [12,13], the dynamical coupling between mechanical vibrations and light (i.e. optomechanical coupling) remains as an open subject. In particular, the presence of strong excitonic resonances in monolayer TMD semiconductors [14,15], which strongly interact with light [16,17,18] (can reflect ~ 90 % of the incident light [19,20,21,22,23]), presents a very interesting platform to explore optomechanical effects in atomically thin materials without the need of an optical cavity. Although such cavityless optomechanical coupling has been demonstrated in quantum well heterostructures [24], nanowires [25,26,27] and nanotubes [28], the unique mechanical properties and the extremely strong excitonic effects in monolayer TMDs can enable fundamental studies and applications in new regimes (for instance, regimes with simultaneously strong optical and mechanical nonlinearity).

We demonstrate in this work gate-tunable exciton-optomechanical coupling in suspended monolayer MoSe$_2$. A schematic device structure is shown in Fig. 1a. The silicon back gate can induce charges in the MoSe$_2$ membrane and pull it down, which breaks the mirror symmetry of the mechanical device. Mechanical vibration of the membrane then creates an oscillating strain in the sample, which in turn produces a dynamical spectral shift in the exciton resonance due to the linear strain dependence of the exciton resonance energy [29,30,31]. As a result, illumination of the sample with light near the exciton resonance can generate a radiation force with dynamical backaction [32,33,34], producing exciton-optomechanical effects that we will investigate below. In the following, we will present results at 4 K unless otherwise specified. (See Supporting Information for details of device fabrications.)

Figure 1b shows the sample reflection contrast spectra at varying gate voltages ($V_g = 0 - 40$ V). (Results on negative $V_g$ are similar.) A strong dip in the reflection contrast due to the neutral exciton resonance in monolayer MoSe$_2$ is clearly observed. The sharp effective exciton linewidth ($\gamma \sim 5$ meV) and the large on-resonance reflection contrast (~ 90 %) are consistent with reported results on nearly perfect monolayer mirrors [19,20]. The exciton resonance redshifts with increasing gate voltage due to the build up of an in-plane strain in the sample. The nonlinear gate-induced energy shift is caused by the nonlinear dependence of the sample strain on $V_g$. The theoretical gate voltage dependence of the exciton resonance energy $E_X$ is also plotted in Fig. 1b, which shows good agreement with experiment (Supporting Information).

To characterize the mechanical resonance of our suspended sample, we drive the MoSe$_2$ membrane by a small AC gate voltage ($\delta V_g = 200$ mV peak-to-peak value) superimposed on top of a variable DC gate voltage $V_g$. We illuminate the sample by a laser beam with wavelength (765 nm) near the exciton resonance, and detect the reflectivity change as a function of the driving frequency. (See Supporting Information for details of measurements.) Figure 1c shows the gate voltage dependence of the fundamental mechanical mode in a contour plot. With increasing $V_g$, the resonance frequency first decreases slightly and then increases. The behavior is consistent with results reported in the literature [4,5]. The dependence can be explained by gate modulation of the effective spring constant of the device as illustrated by the solid line fit (Supporting Information). At small $V_g$, our mechanical resonator has high quality factor Q ~ 20,000. The quality factor decreases quickly with increasing $V_g$ due to increased mechanical loss, consistent with results in the literature [5,35].

To demonstrate exciton-mechanical coupling, we investigate the spectral dependence of the mechanical response. To do this we vary the incident photon energy near the exciton resonance and keep a small incident power at $P = 1$ µW in order to minimize perturbations on the device. Figure 2a shows the fractional change in reflectivity $\frac{dR}{R}$ (peak-to-peak value) near the fundamental mechanical mode at varying incident photon energy ($V_g = 40$ V). The peak response as a function of the photon energy detuning $\Delta E \equiv E - E_X$ from the exciton resonance energy $E_X$ is shown in Fig. 2b. The response changes sign at the exciton resonance, maximizes in amplitude at a small detuning ($\Delta E \sim 2.2$ meV), and diminishes with further increase in detuning. The vanishing response at the exciton resonance provides an accurate measure of the position of zero detuning $\Delta E = 0$.

The spectral dependence in Fig. 2b can be understood in terms of a dynamical shift in the exciton resonance (with peak-to-peak amplitude $\delta E_X$) induced by an oscillating strain in the membrane (illustrated in the inset). This produces a fractional change in reflectivity $\frac{dR}{R} = \frac{1}{R}\frac{dR}{dE}\delta E_X$ at the mechanical resonance. The quantity $\frac{1}{R}\frac{dR}{dE}$ can be calculated from the measured reflection contrast spectrum in Fig. 1b. By matching the calculated $\frac{1}{R}\frac{dR}{dE}\delta E_X$ (red curve) to the measured $\frac{dR}{R}$ (data points) in Fig. 2b and treating $\delta E_X$ as a variable, we can estimate $\delta E_X \approx 67$ µeV at $V_g = 40$ V. By performing the same analysis at each $V_g$, we obtain the gate voltage dependence of $\delta E_X$ in Fig. 2c (data points). The result can be compared to the expected $V_g$-dependence of $\delta E_X = \frac{dE_X}{dV_g}\delta V_g$ (red curve) with $\delta V_g = 200$ mV and $\frac{dE_X}{dV_g}$ obtained from the theoretical $V_g$-dependence of $E_X$ in Fig. 1b. Reasonable agreement is seen. The discrepancy is largely due to the uncertainty in determining the absolute value of $\delta E_X$ from the relation $\frac{dR}{R} = \frac{1}{R}\frac{dR}{dE}\delta E_X$. The result confirms the picture of dynamical exciton spectral shift and shows a monotonic increase in the exciton-mechanical coupling strength $\propto \frac{dE_X}{dV_g}$ with $V_g$. (The dynamical exciton spectral shift also provides an effective method to detect thermal vibrations at cryogenic temperatures as shown in Supplementary Fig. S1.)

Equipped with a basic understanding on the exciton-mechanical coupling in our device, we now proceed to examine effects due to dynamical backaction. Figure 3a shows the optically detected fundamental mechanical mode at varying incident photon energy ($V_g = 40$ V). Compared to Fig. 2a, the incident power is increased from $P = 1$ μW to $P = 10$ μW in order to amplify effects from dynamical backaction (Supplementary Fig. S2). Asymmetric dependence on the optical detuning is seen. In particular, the mechanical resonance for blue-detuning is significantly broader than the red-detuned counterpart (zero detuning has zero response). The shift in the mechanical resonance is also asymmetric with detuning. These behaviors are summarized by the detuning dependences of the mechanical linewidth and resonance frequency (obtained by Lorentzian fitting to the mechanical resonance) in Fig. 3b and 3d, respectively. For large detuning from the exciton resonance, the fundamental mechanical mode is largely unperturbed by the incident light. Near the exciton resonance, however, the mechanical resonance frequency redshifts and the mechanical linewidth broadens with asymmetric detuning dependences.

The observations can be understood in a model based on photothermal backaction and gate-induced mirror symmetry breaking in the mechanical device. At $V_g = 40$ V, the monolayer MoSe$_2$ membrane is pulled down by the electrostatic force from the back gate, which breaks the out-of-plane mirror symmetry of the device (Fig. 1a). An oscillating strain due to the mechanical vibration produces a dynamical energy shift in the exciton resonance with amplitude $\delta E_X$. The dynamical shift in turn periodically modulates the photothermal force $F_{ph}$ produced by laser illumination near the exciton resonance, whose magnitude and phase depend on the detuning $\Delta E$ from the exciton resonance. This results in a dynamical backaction on the membrane's mechanical vibration and changes its mechanical linewidth $\Gamma$ and resonance frequency $f$ according to [32,33,34]

$$\Gamma \approx \Gamma_0 \left[1 + Q \frac{\Omega_0 \tau}{1+\Omega_0^2 \tau^2} \frac{(\nabla F_{ph})_{as}}{4\pi\sigma}\right],$$
$$f \approx f_0 \left[1 - \frac{1}{1+\Omega_0^2 \tau^2} \frac{(\nabla F_{ph})_{as}}{8\pi\sigma}\right]. \quad (1)$$

Here $\Gamma_0$ and $f_0$ are the unperturbed mechanical linewidth and resonance frequency ($\Omega_0 \equiv 2\pi f_0$), respectively, $\tau$ is the time delay for the photothermal force to respond to the membrane's displacement, and $\sigma$ is the unperturbed stress of the membrane, which is directly proportional to the spring constant. Only the anti-symmetric part (in detuning) of the photothermal force gradient $(\nabla F_{ph})_{as}$ associated with a time delay $\tau$ contributes to dynamical backaction [32,33,34]. The result is (Supporting Information)

$$\frac{(\nabla F_{ph})_{as}}{4\pi\sigma} = \alpha P \frac{(\Delta E)}{[(\Delta E)^2+(\gamma/2)^2]^2} \cdot \Delta E_X. \quad (2)$$

Here the pre-factor $\alpha$ depends on material parameters of MoSe$_2$ such as the Young's modulus, the thermal expansion coefficient and the thermal conductivity etc. The result is linearly proportional to both the incident power $P$ and the DC gate-induced exciton resonance energy shift $\Delta E_X$ (Fig. 1b), and has antisymmetric dependence on the detuning $\Delta E$. (The linear power dependence is shown in Supplementary Fig. S3.)

In addition to the dynamical backaction contribution, there is also a contribution from photothermal softening of the mechanical spring constant $(\nabla F_{ph})_s \propto \frac{P}{(\Delta E)^2 + (\gamma/2)^2}$ (Supporting Information). Since this contribution is originated from optical absorption near the exciton resonance, it shows symmetric dependence on the detuning $\Delta E$ and is independent of $\Delta E_X$. The results in Fig. 3b and 3d can be well described by including the two contributions and treating the overall amplitude and the effective exciton linewidth $\gamma$ as fitting parameters. We have also subtracted the photothermal softening contribution from the results in Fig. 3b and 3d, and isolated the antisymmetric contribution in Fig. 3c and 3e. Clear optical spring effect as well optical damping and antidamping arising from dynamical backaction are observed.

The observed dynamical backaction originated from exciton-optomechanical coupling is similar to that in cavity optomechanics [33,34]. In cavity optomechanics, the mechanical vibrations of the end mirror of an optical cavity dynamically change the cavity resonance frequency, which in turn periodically modulates the intracavity optical field and the radiation pressure on the mirror with a time delay $\tau$ given by the photon storage time of the cavity. The delayed radiation pressure creates a dynamical backaction on the mirror's mechanical vibrations, and produces the well-known optical spring effect and optical damping [33,34]. In our case the MoSe$_2$ membrane replaces the mirror, the exciton resonance replaces the cavity resonance, and the photothermal force replaces the radiation pressure (which limits the application of the device on quantum state control and quantum information science in general). The photon storage time now becomes the thermal diffusion time of the suspended MoSe$_2$. Moreover, because the exciton resonance redshifts ($\Delta E_X < 0$) with increasing strain (i.e. increasing downward displacement $z$ of the membrane), the sign of the detuning dependences in Fig. 3c and 3e is opposite to that in cavity optomechanics [33,34]. In particular, red-detuning (blue-detuning) causes antidamping (damping) and blueshift (redshift) on the mechanical resonance (Supporting Information).

Finally, to test the dependence of $(\nabla F_{ph})_{as}$ on $\Delta E_X$ as predicted by Eqn. (2), we perform optical detuning studies (as in Fig. 3) at different gate voltages, which continuously tune the DC exciton resonance shift $\Delta E_X$ (Fig. 1b). We only focus on the mechanical linewidth because optical damping is much more pronounced than the optical spring effect due to the enhancement factor $Q$ in Eqn. (1). Figure 4a shows the detuning dependences of the antisymmetric contribution of the mechanical linewidth at varying $V_g$ (the incident power is fixed at $P = 3.3$ μW). We have subtracted the symmetric contributions from the total (symmetric plus antisymmetric) mechanical linewidth to obtain Fig. 4a. We can see that the amplitude of the antisymmetric contribution (half of the peak-to-peak height in the detuning dependences) increases quickly with $V_g$ (Fig. 4b). We compare the result with the prediction from Eqn. (1) and (2) using the gate-induced $\Delta E_X$ extracted from Fig. 1b and the gate-dependent $\Omega_0$ and $Q$ from Fig. 1c (Method). The agreement further supports our interpretation based on exciton-optomechanical backaction. The result also demonstrates gate-tunable exciton-optomechanical coupling strength.

In conclusion, we have demonstrated dynamical light control of the mechanical motion of suspended monolayer $MoSe_2$ through the material's strong excitonic resonance. Gate-tunable optomechanical damping and antidamping consistent with the picture of photothermal backaction have been observed. Further optimization of the device that allows larger gate-induced exciton resonance shift $\Delta E_X$, which favors dynamical backaction, could open up applications in optomechanical cooling [36] and mechanical lasing [37] in monolayer TMD semiconductors.

The Supporting Information is available free of charge on the ACS Publications website at DOI: 10.1021/acs.nanolett.0c05089.
**It details methods for device fabrication, optical measurements, gate-dependent data analysis, photothermal backaction modeling and supplementary figures.

**Acknowledgements**
This work was primarily supported by the Air Force Office of Scientific Research under Award No. FA9550-18-1-0480. The growth of $MoSe_2$ crystals was supported by the US Department of Energy (DOE), Office of Science, Basic Energy Sciences (BES), under award number DE-SC0019481. Device fabrication was performed in part at the Cornell NanoScale Facility, a member of the National Nanotechnology Coordinated Infrastructure (NNCI), which is supported by the NSF (Grant No. NNCI-1542081). K.F.M. acknowledges support from the David and Lucille Packard Fellowship.

**Figures and figure captions**

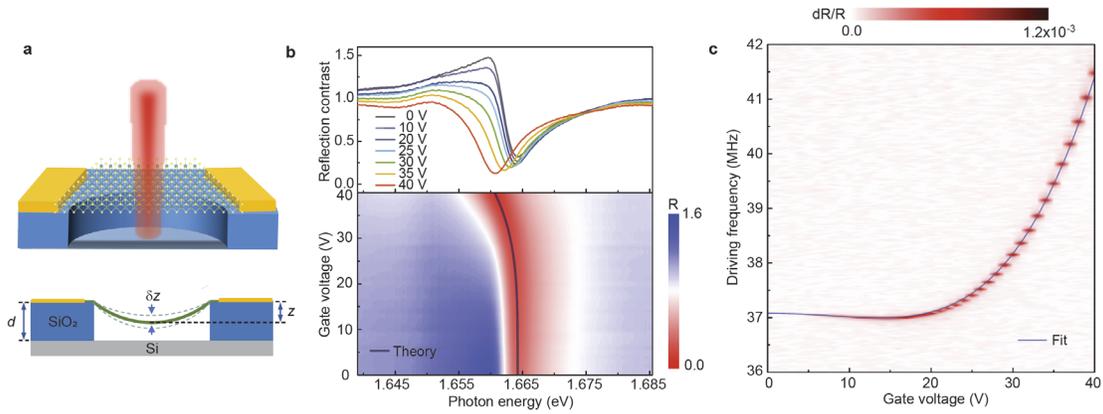

**Figure 1. Monolayer MoSe$_2$ optomechanical device. a,** Schematic illustration of suspended monolayer MoSe$_2$ contacted by gold electrodes and illuminated by a laser beam. *d*: circular trench depth; *z*: static vertical displacement of the membrane induced by a DC gate voltage; *δz*: vibration amplitude induced by an AC gate voltage, which induces a dynamical shift in the exciton resonance and gives rise to dynamical backaction. **b,** Gate voltage dependent reflection contrast spectrum of suspended monolayer MoSe$_2$. Spectra at selected gate voltages are shown in the top panel. The solid line in the contour plot (bottom panel) is the expected dependence of the exciton resonance energy (Supporting Information). **c,** Gate voltage dependence of the mechanical resonance frequency. The solid line is a theoretical fit to the data (Supporting Information).

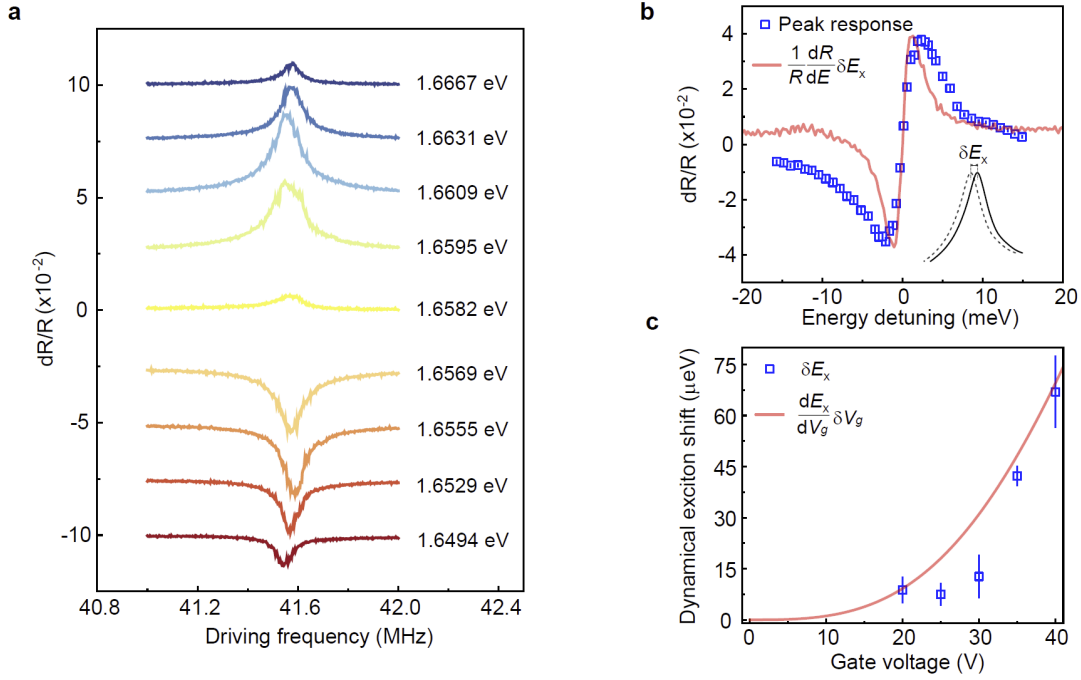

**Figure 2. Dynamical shift in exciton resonance. a,** Fractional change in reflection contrast as a function of the AC driving frequency at different probe photon energies near the exciton resonance (incident power ~ 1 μW). The curves are vertically displaced for clarity purpose. Near zero change is seen at the excitonic resonance. **b,** The peak response extracted from **a** (data point) as a function of the energy detuning from the exciton resonance. The red line is the energy derivative of the normalized reflection contrast spectrum at $V_g = 40$ V (extracted from **Fig. 1b**) multiplied by a dynamical exciton shift with peak-to-peak amplitude $\delta E_X \approx 67$ μeV. Inset: Schematic illustration of dynamical shift in the exciton resonance. **c,** The gate voltage dependence of $\delta E_X$ extracted from similar measurement as in **b** at different gate voltages. The red curve is the expected dependence $\frac{dE_X}{dV_g}\delta V_g$ extracted from the solid line in **Fig. 1b** using a peak-to-peak gate voltage amplitude $\delta V_g = 200$ mV.

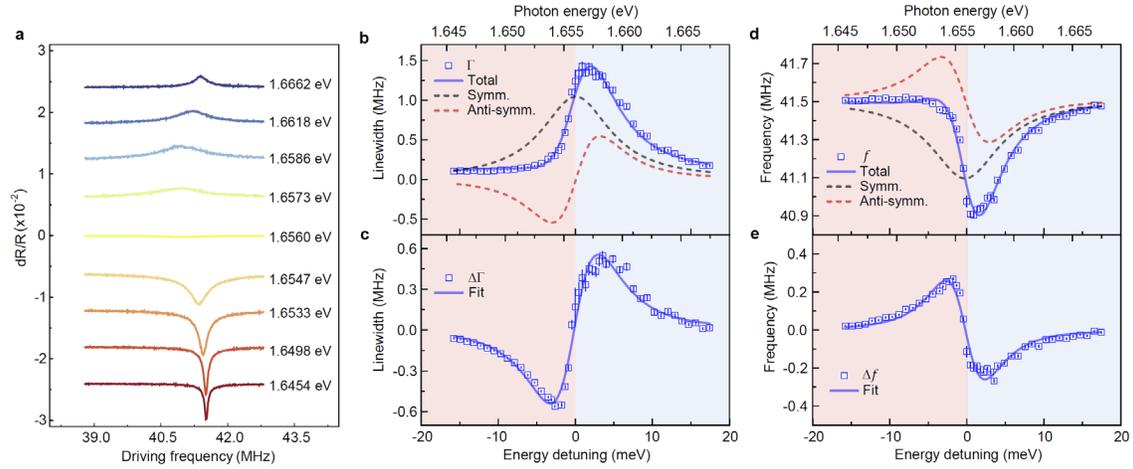

**Figure 3. Dynamical backaction effects. a,** Fractional change in reflection contrast as a function of the AC driving frequency at different incident photon energies near the exciton resonance (incident power ~ 10 µW). The curves are vertically displaced for clarity purpose. **b-e,** The mechanical linewidth (**b, c**) and resonance frequency (**d, e**) extracted from **a** as a function of the energy detuning (bottom axis) and incident photon energy (top axis). The blue and red shaded regions correspond to blue and red energy detuning, respectively. The solid lines in **b** and **d** are fits to the data including both the symmetric photothermal softening contribution (black dashed line) and the anti-symmetric dynamical backaction contribution (red dashed line). Only the anti-symmetric contribution is shown in **c** and **e**, where the symmetric contribution is subtracted off from the data in **b** and **d**.

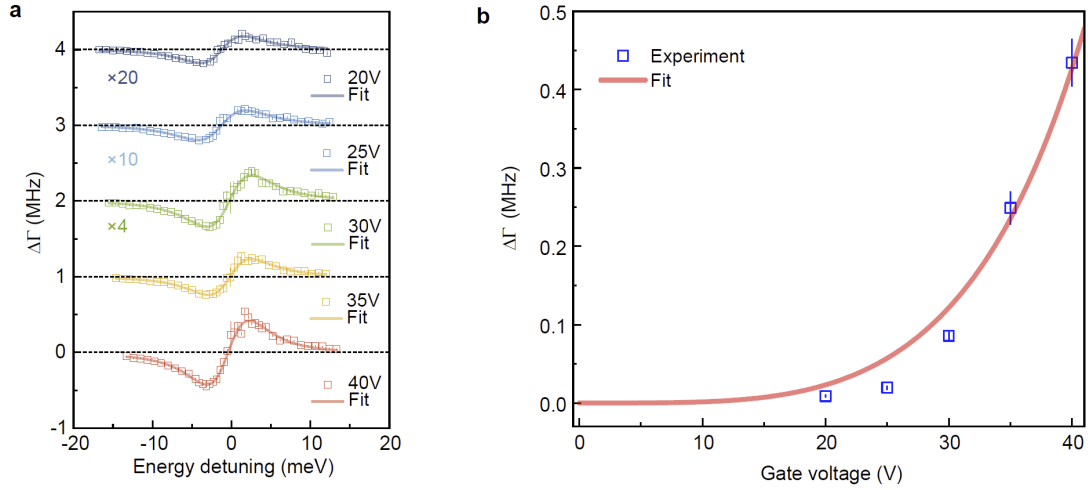

**Figure 4. Gate tunable exciton-optomechanical coupling. a,** The backaction-induced change in mechanical linewidth as a function of the energy detuning at varying gate voltages. The solid lines are fits to the experimental data points using Eqn. (1) and (2). The data are vertically displaced and scaled by different amplitudes for clarity purpose. The dashed lines mark the position of zero change in linewidth. **b,** Half of the peak-to-peak height extracted from **a** as a function of gate voltage. The red curve is the fit to the data using Eqn. (1) and (2) (details see Supporting Information).

TOC Graphic

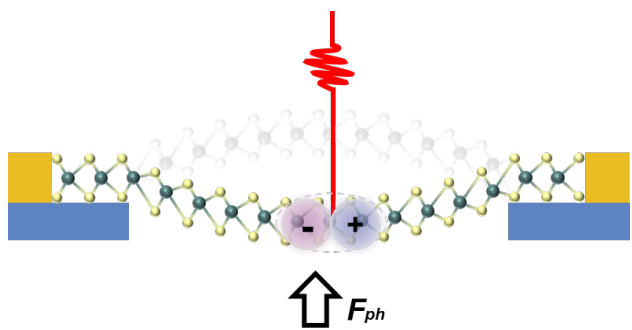

Supporting Information for

# Tunable exciton-optomechanical coupling in suspended monolayer MoSe$_2$

Hongchao Xie[1,2], Shengwei Jiang[1], Daniel A. Rhodes[3], James C. Hone[3], Jie Shan[1,4*], and Kin Fai Mak[1,4*]

[1] Laboratory of Atomic and Solid State Physics and School of Applied and Engineering Physics, Cornell University, Ithaca, New York, USA
[2] Department of Physics, Penn State University, University Park, Pennsylvania, USA
[3] Department of Mechanical Engineering, Columbia University, New York, USA
[4] Kavli Institute at Cornell for Nanoscale Science, Ithaca, New York, USA

*E-mails: jie.shan@cornell.edu; kinfai.mak@cornell.edu

**Content**
1. **Device fabrication**
2. **Reflection contrast measurements**
3. **Measurements of the mechanical resonance**
4. **Gate-induced exciton energy shift**
5. **Gate dependent mechanical resonance frequency**
6. **Photothermal force gradient**
7. **Supplementary figures**
8. **References**

## 1. Device fabrication

To fabricate suspended monolayer MoSe$_2$ resonators, we first patterned circular trenches (~ 600 nm in depth) on SiO$_2$/Si substrate using photolithography and reactive ion etching. Metal pads (Ti 5 nm/Au 30 nm) were subsequently deposited next to the circular trenches. MoSe$_2$ monolayers were mechanically exfoliated from bulk crystals and transferred onto the targeted trench using a polydimethylsiloxane (PDMS) stamp. The monolayers were first identified by optical contrast and further confirmed by reflection contrast and photoluminescence measurements.

## 2. Reflection contrast measurements

We focused broadband radiation from a supercontinuum light source onto the suspended sample using a high numerical aperture (N.A. = 0.8) objective. The beam spot is about 1$\mu$m in diameter and the incident power is below 1 $\mu$W to minimize heating effects. The reflected light is collected by the same objective and detected by a spectrometer equipped with a liquid nitrogen-cooled charge coupled device (CCD). The reflection contrast spectra were obtained by normalizing the reflected light intensity from the suspended MoSe$_2$ region to that from a bare circular trench without MoSe$_2$. The suspended sample was mounted inside a close-cycle cryostat filled with helium exchange gas (< 10$^{-5}$ Torr) at a base temperature (~ 4 K).

## 3. Measurements of the mechanical resonance

We measure the out-of-plane displacement of the resonator by optical interferometry. The optical setup is the same as the reflection contrast measurement. The output of a wavelength-tunable continuous-wave Ti-Sapphire laser was employed for both excitation and probe. We vary the wavelength near the exciton resonance of MoSe$_2$. The reflected beam was collected by the same objective to a fast photodiode. Motion of the membrane was actuated capacitively by a small AC voltage (peak-to-peak amplitude = 200 mV) between the sample and the gate in the linear response regime. The vibration of the membrane, which modulates the reflected light intensity at the driving frequency, was measured by a network analyzer (Agilent E5061A). The DC reflected light intensity was also measured. The fractional change in reflectance $\frac{dR}{R}$ can be obtained by normalizing the AC signal to the DC signal. In thermal vibration measurements (Supplementary Fig. S1), there was no AC driving voltage and the noise voltage density of the photodiode output was measured by a spectrum analyzer (Agilent E4408B).

Note that when the sample is illuminated by the Ti:Sapphire laser for mechanical measurements, the exciton resonance of the membrane is red-shifted because of laser heating (e.g. compare the exciton resonance energy in Fig. 2 with that in Fig. 1b). For a constant incident laser power, however, we can offset this redshift by studying the dependence of the mechanical response on the energy detuning $\Delta E$ instead of the absolute photon energy $E$. The approximately constant redshift for a given incident power has negligible effect on our analysis.

### 4. Gate-induced exciton energy shift

The exciton resonance energy of monolayer MoSe2 is known to redshift linearly with strain [38,39]. We obtain the gate dependent strain $\epsilon(V_g) \approx \frac{1}{96}\left(\frac{R}{d}\right)^2 \left(\frac{\varepsilon_0}{d\sigma_0}\right)^2 V_g^4$ (valid when it is small compared to the built-in strain) by minimizing the total energy $\frac{3}{2}Y\epsilon^2 - \frac{1}{2}C_g V_g^2$ of the device with respect to strain $\epsilon$. Here $R$ is the drumhead radius, $d$ is the trench depth, $\varepsilon_0$ is the vacuum permittivity, $\sigma_0$ is the built-in stress, $Y$ is the Young's modulus and $C_g$ is a $V_g$-dependent geometrical gate capacitance. The gate-induced exciton energy shift in Fig. 1b can be fitted by $\Delta E_X \approx \epsilon(V_g) \times 55$ meV/%. The fitted slope 55 meV/% is consistent with the reported value ~ 54 meV/% in the literature [Ref. 38,39,[40]].

### 5. Gate dependent mechanical resonance frequency

The mechanical resonance frequency for a fully clamped drumhead is given by $f = \frac{2.405}{2\pi R}\sqrt{\frac{\sigma}{\rho}}$ with stress $\sigma = \frac{R^2}{4}\frac{\partial^2}{\partial z^2}\left[\frac{3}{2}Y\epsilon^2 - \frac{1}{2}C_g V_g^2\right]$ [Ref.[41],[42],[43]]. The extracted parameters for our MoSe2 resonator by fitting this expression to the experimental gate dependent frequency in Fig. 1c are $Y = 170 \pm 20$ Nm$^{-1}$, $\rho = (9.0 \pm 0.7) \times 10^{-6}$ kgm$^{-2}$ and $\sigma_0 = 0.53 \pm 0.04$ Nm$^{-1}$. The built-in stress $\sigma_0$ obtained here is consistent with the value $0.43 \pm 0.07$ Nm$^{-1}$ extracted independently from the exciton energy shift discussed above. The 2D mass density $\rho$ is about twice of the expected value $4.62 \times 10^{-6}$ kgm$^{-2}$ presumably due to the presence of PDMS residue on the membrane. The Young's modulus $Y$ is also consistent with the reported values $124\pm7$ Nm$^{-1}$ [Ref. [44]] and $157\pm29$ Nm$^{-1}$ [Ref. [45]] in the literature.

### 6. Photothermal force gradient

Here we derive an expression for the photothermal force gradient $\nabla F_{ph}$, which has been used in the analysis in Fig. 3 and Fig. 4. Based on Ref. 46, the photothermal force gradient can be written in terms of the photothermal contribution to the stress $\sigma_{ph}$ as

$$\nabla F_{ph} = 4\pi \left(\sigma_{ph} + z\frac{d\sigma_{ph}}{dz}\right). \tag{S1}$$

The first term corresponds to photothermal softening of the spring constant and does not contribute to dynamical backaction effects. The second term is the contribution related to dynamical backaction. The photothermal contribution to the stress can be written as [46]

$$\sigma_{ph} = -Y\frac{\alpha_{th}}{4\pi\kappa} L \cdot A \cdot P. \tag{S2}$$

Here $\alpha_{th}$ is the thermal expansion coefficient, $\kappa$ is the sheet thermal conductance, $L$ is local electric field factor for light and $P$ is the incident laser power. The local field factor $L$ is determined by matching the boundary conditions for the electric and magnetic fields at the silicon back gate interface and at the sample plane [47]. It takes into account the multiple interference effects of the device near the exciton resonance. Its photon energy dependence is shown in Supplementary Fig. S4. The absorbance $A = \frac{A'_0(\gamma'/2)^2}{[E-E_X(z)]^2+(\gamma'/2)^2}$ of monolayer MoSe$_2$ is modeled by a Lorentzian representing the exciton resonance ($A'_0$ is the peak absorbance). The exciton resonance energy $E_X(z)$ is z-dependent. The total absorbance, taking into account the local field corrections from multiple interference, is $L \cdot A$, which can be fitted by a Lorentzian $L \cdot A = \frac{A_0(\gamma/2)^2}{[E-E_X(z)]^2+(\gamma/2)^2}$ with a broadened effective exciton linewidth $\gamma > \gamma'$ (Supplementary Fig. S4). ($A_0$ is the total peak absorbance.) We then obtain

$$\left(\nabla F_{ph}\right)_{as} = 4\pi z \frac{d\sigma_{ph}}{dz} \approx -Y \frac{\alpha_{th}}{\kappa} A_0 \cdot P \frac{\Delta E \gamma^2 /2}{[(\Delta E)^2+(\gamma/2)^2]^2} \cdot \frac{dE_X}{dz} z. \tag{S3}$$

Here $\Delta E = E - E_X$ is the photon energy detuning. The strain-induced shift in the exciton resonance is $\Delta E_X \approx \frac{dE_X}{d\epsilon}\left(\frac{2z^2}{3R^2}\right) = \frac{1}{2}\frac{dE_X}{dz} z$. We therefore obtain Eqn. (2) in the main text with the pre-factor equal to $\alpha = -\frac{Y}{\pi\sigma_0}\frac{\alpha_{th}}{\kappa}A_0\left(\frac{\gamma}{2}\right)^2$. To describe the data in Fig. 4b, we use Eqn. (1) for $\Gamma$ and treat $\alpha$ as a fitting parameter. We obtain a good fit using $\alpha \approx 4.6 \times 10^{-20}$ eVs. This value can be compared to the expected value of $\alpha \approx 3.3 \times 10^{-20}$ eVs using $\alpha_{th} \sim 1 \times 10^{-6}$ K$^{-1}$ [Ref. [48]], $\kappa \sim 1 \times 10^{-8}$ WK$^{-1}$ [Ref. [49,50,51]], $A_0 \approx 0.77$, $\gamma \approx 10$ meV and $\tau \approx 3$ns [Ref. 50]. Note that we have used the values of $\alpha_{th}$ and $\kappa$ near 20 K due to laser heating.

Note that the sign of $\left(\nabla F_{ph}\right)_{as}$ is opposite to that in cavity optomechanics based on radiation pressure. This gives rise to the observed antidamping (damping) of the mechanical resonance for red-detuning (blue-detuning) from the exciton resonance. This peculiar dependence, which has also been observed in Ref. [52], can be understood in the following picture. In a vibration cycle, as the membrane moves downward (corresponding to a positive vertical displacement) by responding to a positive driving force from the AC gate voltage, the sample strain increases and induces an exciton redshift. This results in an increase in the optical absorption for red detuning, and in turn softens the membrane and reduces the restoring force. The reduced restoring force (a negative contribution to the spring constant) is equivalent to an extra driving force in phase with the external driving force from the gate. The net result from such dynamical feedback is therefore anti-damping (or heating) of the mechanical vibration. The sign of the effect is reversed for blue detuning.

## 7. Supplementary figures

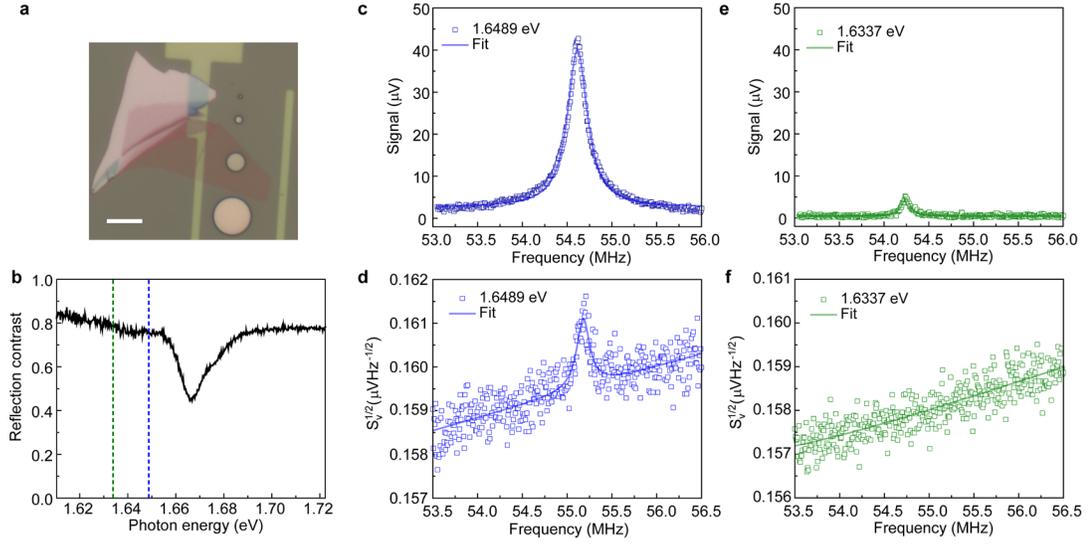

**Supplementary Figure S1. Detection of thermal vibrations at 10 K. a,** Optical image of a typical suspended monolayer MoSe$_2$ device. The scale bar is 10 μm. **b,** The reflection contrast spectrum at $V_g = 60 V$ showing an exciton resonance at 1.666 eV. Dashed lines mark the two probe photon energies (1.6489 eV and 1.6337 eV) employed in **c-f**. Both of them are lower than the exciton resonance energy. **c, e,** Reflected light signal as a function of the driving frequency at 1.6489 eV (**c**) and 1.6337 eV (**e**) probe. The peak-to-peak driving amplitude is 40 mV, the incident power is 10 μW, and the measurement temperature is 4 K. The signal at the probe energy near the exciton resonance is enhanced. **d, f,** Noise voltage density as a function of frequency at 1.6489 eV (**d**) and 1.6337 eV (**f**) probe. The probe power here is 100 μW and there is no driving voltage. The thermal vibration at 10 K at the mechanical resonance is only observed at the probe photon energy near the exciton resonance. Because the probe photon energy is substantially below the exciton resonance energy, there is minimal heating effect in these measurements.

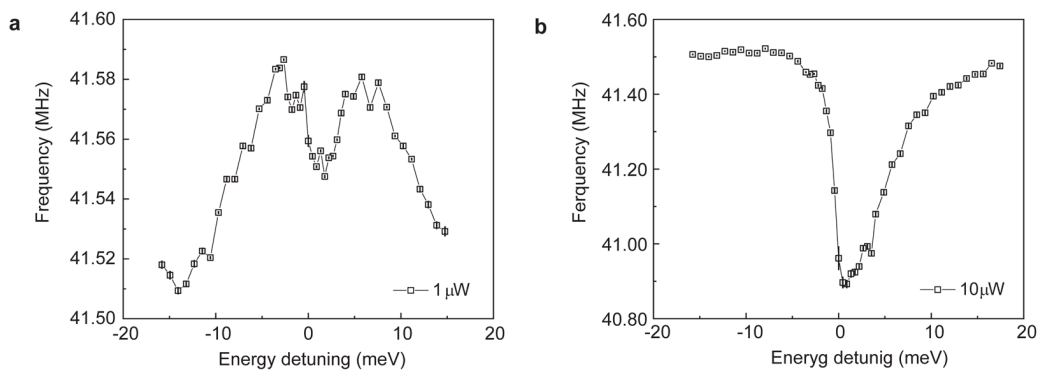

**Supplementary Figure S2. Dependence of the mechanical resonance frequency on energy detuning at 1 µW (a) and 10 µW (b) incident power ($V_g = 40V$).** In addition to the modification of the mechanical resonance frequency near the exciton resonance, there is also a smooth background frequency shift at 1 µW incident power. At 10 µW incident power, the smooth background becomes negligible and only the mechanical frequency shift near the exciton resonance is observed. The origin of the smooth background is unclear. It has much weaker power dependence compared to the photothermal contribution. Excluding this smooth background, the power dependence of the mechanical frequency shift at the exciton resonance is linear and is consistent with Eqn. (S2) and (S3).

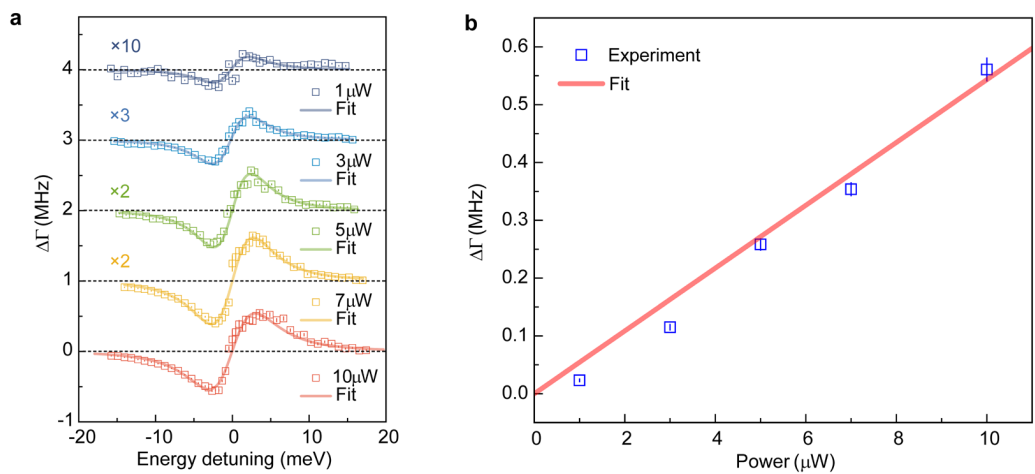

**Supplementary Figure S3. Power dependence of optical damping. a,** The backaction-induced change in mechanical linewidth as a function of the energy detuning at varying incident power. The gate voltage is fixed at 40 V. The solid lines are fits to the experimental data points using Eqn. (1) and (2). The data are vertically displaced and scaled by different amplitudes for clarity purpose. The dashed lines mark the position of zero change in linewidth. **b,** Half of the peak-to-peak height extracted from **a** as a function of incident power. A linear dependence is seen as expected.

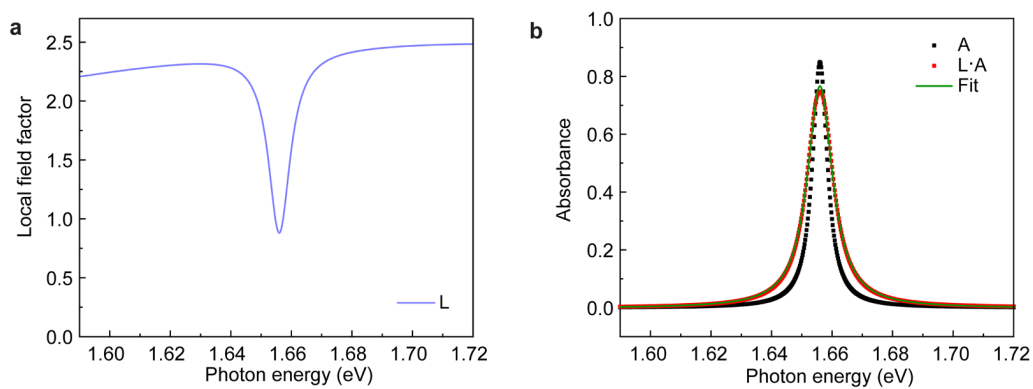

**Supplementary Figure S4. Effects of multiple interference near the exciton resonance.**
**a,** Dependence of the local field factor *L* on photon energy. A dip at the exciton resonance is seen. **b,** Dependence of the intrinsic absorbance *A* of the sample (black) and the total absorbance $L \cdot A$ including local field effects (red) on photon energy. Similar to the intrinsic absorbance, the total absorbance can be well fitted by a Lorentzian (green). The resultant effective exciton linewidth is broader than the intrinsic exciton linewidth due to local field corrections.